\documentclass{aa}
\usepackage{natbib}
\usepackage{psfig}
\usepackage{txfonts}

\def\eps{\varepsilon}

\makeglossary
\def\glo#1#2{\item[#1 :] #2}

\def\nat{Nature }
\def\mnras{MNRAS }
\def\aap{A\&A }
\def\aj{AJ }
\def\apj{ApJ }
\def\apjl{ApJL }
\def\araa{ARAA }
\def\solphys{Sol. Phys. }

\newlength{\figwidth}
\def\figwidth{0.5 \textwidth}

\begin{document}

\title{On the escape of cosmic rays from radio galaxy cocoons}
\titlerunning{On the escape of cosmic rays}
\author{Torsten A.  En{\ss}lin}
\authorrunning{T.A.  En{\ss}lin}
\institute{Max-Planck-Institut f\"{u}r Astrophysik,
Karl-Schwarzschild-Str.1, 85740 Garching, Germany\\Department of
Physics, University of Toronto, 60 St. George Street, Toronto, M5S1A7,
Canada} 
\date{Received 2 November 2000 / Accepted 12 November 2002}
\abstract{The escape rate of cosmic ray (CR) particles from radio
galaxy cocoons is a problem of high astrophysical relevance: e.g. if
CR electrons are stored for long times in the dilute relativistic
medium filling the radio cocoons ({\it radio plasma} in the following)
they are protected against Coulomb losses and thus are able to produce
a significant non-thermal Comptonisation signature on the CMB. On the
other hand, CR protons and positrons which leak out of radio plasma
can interact with the ambient medium, leading to characteristic gamma
ray radiation by pion decay and pair annihilation.  In order better
understand such problems a model for the escape of CR particles from
radio galaxy cocoons is presented here. It is assumed that the radio
cocoon is poorly magnetically connected to the environment. An extreme
case of this kind is an insulating boundary layer of magnetic fields,
which can efficiently suppress particle escape. More likely, magnetic
field lines are less organised and allow the transport of CR particles
from the source interior to the surface region. For such a scenario
two transport regimes are analysed: diffusion of particles along
inter-phase magnetic flux tubes (leaving the cocoon) and cross field
transport of particles in flux tubes touching the cocoon surface.  The
cross field diffusion is likely the dominate escape path, unless a
significant fraction of the surface is magnetically connected to the
environment.  Major cluster merger should strongly enhance the
particle escape by two complementary mechanisms. i) The merger shock
waves shred radio cocoons into filamentary structures, allowing the
CRs to easily reach the radio plasma boundary due to the changed
morphology. ii) Also efficient particle losses can be expected for
radio cocoons not compressed in shock waves. There, for a short period
after the sudden injection of large scale turbulence, the (anomalous)
cross field diffusion can be enhanced by several orders of
magnitude. This lasts until the turbulent energy cascade has reached
the microscopic scales, which determine the value of the microscopic
diffusion coefficients.  }
\maketitle

%\tableofcontents

\section{Introduction}
\glossary{\glo{CR}{cosmic ray}}
\glossary{\glo{CMB}{Cosmic Microwave Background}}

\subsection{Motivation\label{sec:motiv}}
The outflows of radio galaxies fill large regions in the intergalactic
space with relativistic, magnetised plasma. The fate of this radio
plasma is unclear, since it rapidly becomes undetectable for radio
telescopes due to the radiative energy losses of the higher energy
electrons, which emit the observable synchrotron emission at radio
bands. Lower energy electrons, and any possible present relativistic
proton population, may reside for cosmological times, unless they are
able to escape by spatial diffusion.

The escape of CRs out of radio plasma is of high astrophysical
relevance. If CR electrons are able to leak out of the cocoon, they
are possible seed particles for the Mpc-sized radio halos in clusters
of galaxies if they are re-accelerated in the cluster turbulence to
radio-observable energies
\citep{1993ApJ...406..399G,2001MNRAS.320..365B}. If radio cocoons
release CR protons into the intra-cluster medium, secondary CR
electrons are produced in hadronic interactions with the background
gas, which also could explain the existence of cluster radio halos
\citep{1980ApJ...239L..93D}. If radio plasma contains a significant
fraction of positrons, they could lead to a detectable annihilation
line, if they would be able to leave the radio cocoon and interact
with a dense ambient intra-cluster medium \citep{2002ApJ...572..796F}.

On the other hand, if the relativistic electrons are efficiently
confined for cosmological time-scales in radio cocoons, they are
shielded from severe energy losses by Coulomb interaction with the
environmental gas. In such a case they would be long-lived and would
therefore be able to produce a non-thermal Comptonisation
signature in the cosmic microwave background \citep{ensslin2000a,
2002A&A...383..423E}. Further, if old, invisible radio plasma with
confined relativistic electrons is dragged into a shock wave of the
large-scale structure formation, its radio emission is possibly
revived and forms the observed {\it cluster radio relics}
\citep[and references therein]{2002MNRAS.331.1011E}.

The recent detections of several ghost cavities in galaxy clusters
\citep[for lists of recent detections e.g. ][]{2002A&A...384L..27E,
2002ApJ...573..533S}, which are often but not always radio-emitting,
support the assumption that the environmental gas and the bulk of the
relativistic plasma stays separated on a timescale of 100 Myr. However,
leakage of higher energy particles is not excluded by these
observations.

CR escape from radio plasma can also play a role in our
galaxy. Microquasars are proposed to contribute characteristic
spectral features to the galactic CR spectrum, if CRs are able to
leave the ejected plasma \citep{2002A&A...390..751H}.

\subsection{A simplified approach}
The usual description of CR escape from systems like our Galaxy is
done in terms of {\it leaky box} models, where the CR particles
trapped in some region are attributed a characteristic escape time or
frequency ($\tau = \nu_{\rm escape}^{-1}$). This escape time is often
empirically determined (e.g. for the CR escape from the Galaxy by
radioactive CR clocks \citep[e.g. ][]{1980ARA&A..18..289C}).  It is
the goal of this work to provide an estimate of this escape time and
its dependency on the geometrical and plasma parameters of the system
under investigation. The focus of this work is cocoons of radio
galaxies, but the presented model may have other applications.

In many applications, the physical parameters will not be sufficiently
constrained to allow an accurate calculation of the escape
time. Nevertheless, insight into the dependencies on parameters such
as turbulence, fraction of the magnetically open surface etc. allows
statements of the relative CR escape rates in different situations to
be made. This should help to formulate hypotheses about the conditions
in which CR escape is efficient, which may be tested
observationally. Examples of such tests were given in
Sec. \ref{sec:motiv}.

The following simplifications are used to compile the CR escape
model: 
\begin{itemize}
\item
Convective CR transport, e.g. by plasmoids which detach by
reconnection events from the source region, is neglected.
\item
The detailed magnetic structure of the source interior is not modelled
and is treated as being homogeneous. Only the topological properties of
the magnetic flux tubes touching the source surface are modelled.
\item
The magnetic field at the source region  surface is virtually split into flux
tubes with diameters of the perpendicular autocorrelation length of the
magnetic field. The transport of particles along and across the flux
tubes is described as a diffusion process. A similar phenomenological
description of CR transport within flux tubes can be found in
e.g. \cite*{2000ApJ...529..513C}, where limits of such a picture
are also discussed.
\item
It is assumed that such a surface flux tube bends at some point into
the interior of the source region. This allow CRs from the interior to
enter the tube there and to follow it to the surface.  In the case
that the magnetic fields are more onion-like in structure, the
particle transport will be much more suppressed than predicted with
this model.  A brief discussion of such a hypothetical situation is
given in Sect. \ref{sec:layer}.
\item
A simplified description of the diffusion process is adopted: The
microscopic diffusion coefficients ($\kappa_\|$, $\kappa_\perp$) are
parametrised. The spectrum of magnetic field fluctuations, which
strongly determines the transport coefficients, is assumed to be a
single power law, connecting the length scales on which field line
wandering happens, down to CR gyro-radii scales. In any numerical
example a Kolmogorov-like turbulence spectrum is adopted. The detailed
dependency on the turbulence wave types
\citep[e.g.][]{1997A&A...326..793M, 1998A&A...337..558M,
1999SoPh..184..339M, 2000PhRvL..85.4656C} is put into a phenomenological
``fudge factor''.
\item
The macroscopic cross field diffusion coefficient ($\kappa_{\rm a}$)
is assumed to exceed the microscopic one by orders of magnitude due to
magnetic field line wandering. This allows CRs that are rapidly
diffusing along the field lines to be transported across them
\citep{1978PRL...40..38}. This also ensures that the length scales of
the diffusion problem are always shorter than length scales on which a
magnetic flux tube loses its identity due to field line
wandering. This should justify the use of the flux tube picture.
\end{itemize}

\subsection{The structure of the paper}
The structure of this article is the following: In
Sect. \ref{sec:basics} the theoretical tools for the description of
the CR propagation in inhomogeneous media are compiled. The
inter-phase CR transport is analysed in Sect. \ref{sec:escape}. Three
escape routes are considered: the penetration of an isolating boundary
layer (Sect. \ref{sec:layer}), the parallel diffusion along
inter-phase magnetic flux tubes (Sect. \ref{seq:paralell}), and the
cross field escape (Sect. \ref{seq:perpendicular}). In
Sect. \ref{sec:appl} the CR escape from radio plasma cocoons is
investigated and in Sect. \ref{sec:diss} the main findings of the
paper are listed. In Appendix A the spatially homogenised transport
equation of CRs in a small-scale inhomogeneous medium is derived.
Appendix B contains a glossary of the frequently used symbols and
abbreviations.

\section{Cosmic ray diffusion\label{sec:basics}}
\subsection{Diffusion in inhomogeneous magnetic fields}
Since the magnetic field topology dominates the mobility of CR
particles, adapted coordinates are chosen. The length $x$ is measured
along the (local) mean direction $\vec{B}$ of a field bundle with
local field strength $\vec{M} = \vec{B} + \delta\vec{B}$. Although
individual field lines may leave the flux tubes, due to magnetic
fluctuations $\delta\vec{B}$, this {\it mean field flux tube} is well
defined and gives an appropriate local coordinate system.

The analysis is restricted to particles with gyro-radii much smaller
than the length-scale of the magnetic fields. In this case the cross
field diffusion is orders of magnitude slower than the diffusion along
the field lines.  Particles can be regarded as being confined in a
flux tube. Since the diameter and the field strength of the tube can
change as a function of $x$, it is convenient to work with the number
of particles $F$ per flux tube length $dx$ and magnetic flux $d\phi_B
= B\, dy\,dz$\glossary{\glo{$\phi_B$}{magnetic flux}}, instead of the
usual volume normalisation $dV = dx\,dy\,dz = dx\, d\phi_B/B$.  The
particle phase space distribution function is $F(x,p,\mu,t)\, = d
N/({dx\,d\phi_B\,dp\,d\mu})$, and its source density is
$Q(x,p,\mu,t)\, = d \dot{N}/({dx\,d\phi_B\,dp\,d\mu})$. $\vec{p}$ is
the momentum of the particle, $v$ its velocity, $m$ is its mass, and
$c$ is the speed of light.\glossary{\glo{$p$}{particle
momentum}}\glossary{\glo{$m$}{particle
mass}}\glossary{\glo{$v$}{particle velocity}} $\mu = p_x/p $ is the
cosine of the pitch angle between $\vec{B}$
\glossary{\glo{$\mu$}{cosine of the CR pitch angle}}(or $\vec{x}$) and
$\vec{p}$. The particle distribution is assumed to be rotationally
symmetric with respect to $\vec{B}$, therefore the azimuthal angle has
been integrated out. The particles entering or leaving the flux tube
are included in the source term $Q$ and in the loss term
$-F/\tau$\glossary{\glo{$\tau$}{particle escape time from a flux
tube}} in the Fokker-Planck equation for $F$:
\begin{eqnarray}
&&\frac{\partial F}{\partial t} 
+ \frac{\partial }{\partial x} (v \,\mu\,\, F) + \frac{\partial
}{\partial p} (\dot{p} \,\,F)
+ \frac{\partial }{\partial \mu} \,
(\dot{\mu}\,F ) =\nonumber\\
\label{eq:FokkerPlanck}
&& 
\frac{\partial }{\partial \mu}  D_{\mu\mu} \frac{\partial }{\partial
\mu} F - \frac{F}{\tau}\, + Q\,.
\end{eqnarray}
$D_{\mu\mu}$ is the Fokker-Planck pitch angle diffusion
coefficient\glossary{\glo{$D_{\mu\mu}$}{Fokker-Planck pitch angle
diffusion coefficient}}, which can in principle be calculated from
quasi-linear theory of plasma wave-particle interaction for a given
spectrum of plasma waves \citep[e.g. ][and references in the
latter]{1966ApJ...146..480J, 1967ApJ...149..405J, 1969ApJ...156..445K,
1970ApJ...162.1049H,1975MNRAS.172..557S,2002cra..book.....S}. Further
Fokker-Planck diffusion coefficients, which describe momentum changes
of the particles, can be found in in the literature
\citep[e.g.][]{1989ApJ...336..243S,1989ApJ...336..264S}. Here,
momentum diffusion is ignored due its small impact on spatial
transport processes, and only continuous energy or momentum losses are
included by $\dot{p}$.  

The continuous pitch angle changes can be calculated from the
adiabatic invariants of a particle with charge
$Z$\glossary{\glo{$Z$}{particle charge}} moving in spatially slowly
varying, and temporally constant magnetic fields. Specifically, these
are the linear momentum $p$ and the magnetic flux within a gyro-radius
$\phi_B(r_g) = \pi \, r_g^2 \,B$, where $r_g(x, p,\mu) =
(1-\mu^2)^{1/2} \,p \,c/ (Z\,e\,B(x))$ is the gyro-radius. One gets
\glossary{\glo{$r_g$}{particle gyro-radius}}
\begin{equation}
\dot{\mu} = - \frac{v}{2}\, (1-\mu^2) \, \frac{\partial B}{B\,\partial x} \,.
\end{equation}
\cite*{1995AstL...21..411K} and \cite*{2000ApJ...529..513C}
investigate the weak scattering regime, in which the mean free path of
a CR is large compared to typical length-scales of the inhomogeneous
magnetic fields and CR density scale-length.  Here, a more
conventional standpoint is adopted, by assuming that the particle
distribution is rapidly isotropised by plasma wave
interactions.\footnote{This is supported by the observation that the
pitch angle diffusion coefficient is the fastest of all the
Fokker-Planck coefficients \citep[e.g.][]{2002cra..book.....S}. The
relatively fast pitch angle scattering should help to maintain a
nearly isotropic CR distribution -- as observed in our own galaxy. A
criterion to test if pitch angle scattering is indeed sufficient to
establish an isotropic pitch-angle distribution -- as assumed
throughout this paper -- is given later.}  Therefore the anisotropic
part of the distribution function is small compared to the isotropic
one. The pitch angle integrated distribution and injection
densities\glossary{\glo{$f$}{isotropic particle phase-space density
per magnetic flux (Eq. \ref{eq:fdef})}}\glossary{\glo{$q$}{isotropic
particle phase-space injection rate per magnetic flux
(Eq. \ref{eq:qdef})}} are defined by
\begin{eqnarray}
\label{eq:fdef}
f(x,p,t) &=& \frac{d N}{dx\,d\phi_B\,dp}      =  \int_{-1}^{1} \!\!\!\!\!d\mu\, F(x,p,\mu ,t)\,,\\
\label{eq:qdef}
q(x,p,t) &=& \frac{d \dot{N}}{dx\,d\phi_B\,dp}=  \int_{-1}^{1} \!\!\!\!\!d\mu \,Q(x,p,\mu ,t)\,.
\end{eqnarray}
With the help of a quasi-linear approximation the evolution equation
for $f$ can be derived from Eq. \ref{eq:FokkerPlanck} following the
calculation steps described in Schlickeiser
(1989a)\nocite{1989ApJ...336..243S}:
\begin{equation}
\label{eq:difff}
\frac{\partial f}{\partial t} + \frac{\partial }{\partial p} (\dot{p}
\,\,f) = \frac{\partial }{\partial x} \left( \kappa_\|
\frac{\partial (B\,f)}{B\, \partial x}\right) - \frac{f}{\tau} + q
\end{equation}
This equation describes the diffusive transport of particles along the
flux tube. In the case that small-scale variations of the
coefficients in this equations exist, effective large-scale
coefficients can be estimated, as shown in 
Appendix A. The diffusion coefficient
\glossary{\glo{$\kappa_\|$}{parallel diffusion coefficient
(Eqs. \ref{eq:diffpar1}, \ref{eq:diffpar2}, \ref{eq:diffpar3})}} is 
\begin{equation}
\label{eq:diffpar1}
\kappa_\|(x,p) = \frac{v^2(p)}{8} \,\int_{-1}^{1} \!\!\!\!\!d\mu
\,\frac{(1-\mu^2)^2}{D_{\mu\mu}(x,p,\mu)} 
\end{equation}
and the pitch angle averaged momentum losses are 
\begin{equation}
\dot{p}(x,p) = \frac{1}{2} \int_{-1}^{1} \!\!\!\!\!d\mu\, \dot{p}(x,p,\mu)\,.
\end{equation}
In order to demonstrate that Eq. \ref{eq:difff} is the proper
transport equation the volume density of particles with momentum $p$
is introduced: $g(x,p,t)\, = d N/({dV\,dp}) =
B(x)\,f(x,p,t)$.\glossary{\glo{$g$}{isotropic particle phase space
density per volume}} Eq. \ref{eq:difff} then transforms into
\begin{equation}
\label{eq:diffg}
\frac{\partial g}{\partial t} + \frac{\partial }{\partial p} (\dot{p}
\,\,g) =B\, \frac{\partial }{\partial x} \left(
\frac{\kappa_\|}{B} \frac{\partial g}{ \partial x}\right)  -
\frac{g}{\tau} + B\,q \,.
\end{equation}
This equation is consistent with Eq. 2 of \cite*{1978A&A....70..367C}.
Diffusion of $g$ vanishes for $\partial g/\partial x = 0$. This means
that diffusion tries to approach a state where the space density of
particles with identical momentum is constant, as it should do. Due to
variations in the diameter of the flux tube this can result in a
non-constant $f$ as a function of position.

\subsection{Diffusion coefficients}

The parallel diffusion coefficient depends on the pitch angle
diffusion coefficient according to Eq. \ref{eq:diffpar1}. By defining
the particle-wave scattering frequency $\nu_\mu$
\glossary{\glo{$\nu_\mu$}{particle-wave pitch-angle scattering
frequency}}
\begin{equation}
\nu_\mu^{-1} = \frac{3}{8} \,\int_{-1}^{1} \!\!\!\!\!d\mu
\,\frac{(1-\mu^2)^2}{D_{\mu\mu}(x,p,\mu)} 
\end{equation}
one can write the parallel diffusion coefficient as
\begin{equation}
\label{eq:diffpar2}
\kappa_\|(x,p) =  \frac{\kappa_{\rm Bohm}(x,p)}{\eps(x,p)} \,.
\end{equation}
Here, $\eps(xp)$ is the ratio of scattering frequency $\nu_\mu$ to the
gyro-frequency $\Omega(x,p) = v(p)/r_g(x,p,0)$, and $\kappa_{\rm
Bohm}(x,p) = v(p) \,r_{g}(x,p,0)/3 = v(p)\,p\,c/(3\,Z\,e\,B(x))$ is
the Bohm diffusion coefficient. $\nu_\mu$ can be regarded as the decay
time of the parallel particle velocity autocorrelations, which in the
following is assumed to be identical to the decay time of the
perpendicular velocity autocorrelations. This latter assumption allows
us to write the microscopic perpendicular diffusion coefficient as
\glossary{\glo{$\kappa_\perp$}{perpendicular diffusion coefficient
(Eqs. \ref{eq:kappaperp1}, \ref{eq:kappaperp2})}}
\begin{equation}
\label{eq:kappaperp1}
\kappa_\perp = \frac{\eps}{1+ \eps^2} \,\kappa_{\rm Bohm}\,
\end{equation}
\citep{1997ApJ...485..655B}.

The scattering frequency $\nu_\mu$ depends on details of the
underlying plasma turbulence on scales comparable to the gyro-radius
of the particle. There are two contributions important to this
small-scale turbulence: The first is the Kolmogorov (or Kraichnan)
cascade of large-scale turbulent kinetic energy to smaller
length-scale, and the second is turbulence induced by CR streaming
\citep{1968ApJ...152..987W,1969ApJ...156..303W,1975Natur.258..687S}.
Since the focus of this work is on the transport of poorly connected
regions, the amount of CR streaming is expected to be low. In the
following numerical examples, only external Kolmogorov-like
turbulence is assumed. In other cases, the theory of this article can
still be applied if the appropriate diffusion coefficients are used.

A simplified parameterisation of the scattering frequency is adopted
here. It is assumed that
\begin{equation}
\label{eq:epsfunc}
\eps(p) = \eps_{\rm 0} \,\delta B^2(r_g(p))/ B^2 \,,
\end{equation}
\glossary{\glo{$\eps$}{ratio of scattering frequency to gyro-frequency
(Eq. \ref{eq:epsfunc})}}\glossary{\glo{$\eps_0$}{fudge factor relating
turbulence energy density to pitch angle scattering efficiency
(Eq. \ref{eq:epsfunc})}} where $\eps_{\rm 0}$ is a fudge factor, which
allows one to trace and correct the error made in this simplification,
and
\begin{equation}
\label{eq:magturb}
\delta B^2(l)= \delta_0 \,B^2\, (l/l_B)^{\gamma}\, ,
\end{equation}
is the power in the magnetic fluctuations on scale $l$ (and smaller).
In Kolmogorov-like turbulence $\gamma = 2/3$, which will be used in
numerical examples\glossary{\glo{$\gamma$}{spectral index of magnetic
turbulence spectrum}}. The reference scale $l_B$ is the largest
length scale which contains significant magnetic power (compared to
the power-law given by Eq. \ref{eq:magturb}), so that $\delta B^2(l_B)
= \langle \delta B^2 \rangle$ is the total magnetic fluctuation
power. This length scale is also of the order of the coherence
length of the fields\glossary{\glo{$l_B$}{magnetic turbulence
reference scale (Eq. \ref{eq:magturb})}}.
\glossary{\glo{$\delta_0$}{relative strength of magnetic turbulence at
reference scale (Eq. \ref{eq:magturb})}} Throughout this paper
(with the exception of the enhanced anomalous diffusion discussed in
Sec. \ref{sec:EAD}) it is assumed that a single power-law describes
the magnetic fluctuation spectrum from the scales important for field
line wandering down to the CR gyro-radii.

The parameter $\delta_0$ and $\eps_0$ play an important role for
many of the addressed questions and their expected values should
briefly be discussed here, although our knowledge of these quantities
is still very limited. $\delta_0$ is roughly speaking the ratio of the
magnetic power on the largest length-scale, on which the power-law
spectrum of the inertia range of the turbulence is valid, to the total
magnetic energy density. For a sharply peaked, single power-law
magnetic spectrum it is $\delta_0 \sim 1$. For a broken power-law
spectrum, or a very broad maximum above the inertia-range
length-scales, we expect $\delta_0 \ll 1$.  $\eps_0$ gives the
efficiency of CR scattering per magnetic power on length-scales of the
order of the particle gyro-radius, thus the power on scales which can
resonate with the particle gyro-orbit. For an Alfv\'en wave spectrum
with slab-like geometry (waves-vectors are mainly parallel to the
magnetic main direction) one expects $\eps_0 \sim 1$.\footnote{As
can be found by an order of magnitude estimations of the pitch angle
scattering frequencies as given by e.g. \cite{2002cra..book.....S}}
However, the nature of MHD turbulence might be anisotropic on small
scales in the sense that mainly wave-modes with wave-vectors
perpendicular to the main field are populated
\citep{1994ApJ...432..612S,1997ApJ...485..680G}. In that case a strong
reduction of the scattering frequency can be expected ($\eps_0 \ll 1$)
\citep[a quasi-linear estimate is provided by][ however it is also
noted there that the quasi-linear approximation is not fully
applicable in this case]{2000PhRvL..85.4656C}.

In order that the propagation of CRs is diffusive the pitch angle
distribution should be sufficiently isotropic. This is given if the
scattering length $l_{\rm scatt}(p) = v/\nu_\mu$
\glossary{\glo{$l_{\rm scatt}$}{CR pitch-angle scattering length
$l_{\rm scatt} = v/\nu_\mu$}} is small compared to any CR density
scale-length ($f/(\partial f/\partial x)$). Within the above
parametrisation we get
\begin{eqnarray}
l_{\rm scatt}(p) &=& \frac{l_B}{\eps_0\,\delta_0} \left(\frac{r_{\rm
g}}{l_B} \right)^{1-\gamma}\\
& =& \frac{10^{-3}\,{\rm
kpc}}{\eps_0\,\delta_0} \, \left(
\frac{p\,c}{\rm GeV} \right)^{\!\frac{1}{3}}
\left(
\frac{l_B}{{\rm kpc}} \right)^{\!\frac{2}{3}}
\left(
\frac{Z\,B}{\mu {\rm G}} \right)^{\!\!-\frac{1}{3}}.
\nonumber
\end{eqnarray}
This is sufficiently small in most of our cases to allow the
diffusive approximation to be used.

The resulting diffusion coefficients for CR particles are then
\begin{eqnarray}
\label{eq:diffpar3}
\kappa_\|  \!\! &=& \!\! \frac{v\,l_B}{3\,\eps_{\rm 0}\,\delta_0} \, \left(
\frac{p\,c}{Z\,e\,B\,l_B} \right)^{1-\gamma }\\& =&
3.2\cdot 10^{28}\,\frac{\rm cm^2}{\rm s} \frac{v}{c\,\eps_{\rm 0}\,\delta_0} \left(
\frac{p\,c}{\rm GeV} \right)^{\!\frac{1}{3}}
\left(
\frac{l_B}{{\rm kpc}} \right)^{\!\frac{2}{3}}
\left(
\frac{Z\,B}{\mu {\rm G}} \right)^{\!\!-\frac{1}{3}}\!\!\!\!\!\!\!\!\!\!\!\!\!\!\!\!\nonumber\\
\label{eq:kappaperp2}
\kappa_\perp  \!\! &=& \!\! \frac{v\,l_B\,\eps_{\rm 0}\,\delta_0}{3} \, \left(
\frac{p\,c}{Z\,e\,B\,l_B} \right)^{\gamma+1 }\\
 \!\! &=& \!\! 3.5\cdot 10^{16}\,\frac{\rm cm^2}{\rm s} \frac{v\,\eps_{\rm 0}\,\delta_0}{c} \left(
\frac{p\,c}{\rm GeV} \right)^{\!\frac{5}{3}}
\left(
\frac{l_B}{{\rm kpc}} \right)^{\!\!-\frac{2}{3}}
\left(
\frac{Z\,B}{\mu {\rm G}} \right)^{\!\!-\frac{5}{3}}\!\!\!\!\!\!\!\!\!\!\!\!\!\!\!\!\nonumber
\end{eqnarray}
Note that these diffusion coefficients are valid under the assumption
that the turbulence dissipation length is smaller than the
gyro-radius. If this is not the case, a much higher $\kappa_\|$ and a
much lower $\kappa_\perp$ would result.

Cross field diffusion is strongly inhibited. The small mobility of
particles perpendicular to the field lines can be amplified by rapid
parallel diffusion along diverging field lines
\citep{1978PRL...40..38}. A particle's microscopic displacement
from its original field line of the order of the gyro-radius by
microscopic cross-field diffusion grows exponentially while it follows
its new field line.  This can be described in terms of a Liapunov
length $\lambda_{\rm L}$, which depends on the behaviour of the
magnetic field autocorrelation function at small displacements, since
the old and new field lines will be strongly correlated. As soon the
particle is sufficiently far from its original field line, it can be
regarded as de-correlated from it. During the time needed for the
particle to de-correlate from its present field line, the particle is
tied to it and follows its stochastic wandering while diffusing along
it. The stochastic field line wandering can be described as a
diffusion process, with a diffusion coefficient $D_B$ which gives the
perpendicular displacement (squared) per unit length travelled along
the field line. Thus, during a de-correlation time the particle is
doing perpendicular steps of the length given by the field line
wandering times the typical diffusion length along the field line
(during this time). Since this leads to a stochastic displacement, the
combined propagation can be described as a macroscopic diffusion
process, which is called anomalous diffusions.  This anomalous
diffusion allows the CRs to use the usually more powerful large-scale
magnetic fluctuations for their transverse propagation. A
mathematical description of this can be found in
\cite*{1995A&A...302L..21D}, which should be consulted for
details. In their formalism the anomalous diffusion depends strongly
on the parameter \glossary{\glo{$\Lambda$}{Eqs. \ref{eq:Lambda0},
\ref{eq:Lambda}}}
\begin{equation}
\label{eq:Lambda0}
\Lambda = \frac{\delta B^2(l_B)\,\lambda_\|}{\sqrt{2}\,
B^2\,\eps(r_g)\,\lambda_\perp}\,,
\end{equation}
where $\lambda_\|$, $\lambda_\perp$ are the correlation length of the
field fluctuations along and perpendicular to the main field direction
(usually $l_{\rm B} \approx \lambda_\| \ge
\lambda_\perp$).\glossary{\glo{$\lambda_\|$}{parallel correlation
length of the field
fluctuations}}\glossary{\glo{$\lambda_\perp$}{perpendicular
correlation length of the field fluctuations, typical flux tube
diameter}} $\Lambda$ is therefore mainly the ratio of the magnetic
turbulence energy density on the largest scales ($l_B$) to that on the
small gyro-radius scale. For the assumed magnetic turbulence spectrum
(Eq. \ref{eq:magturb}) one gets
\begin{equation}
\label{eq:Lambda}
\Lambda = \frac{\delta_0\,\lambda_\|}{\sqrt{2}\,
\eps(r_g)\,\lambda_\perp} = \frac{\lambda_\|}{\sqrt{2}\,
\eps_0\,\lambda_\perp}\, \left( \frac{p\,c}{Z\,e\,B\,l_B}
\right)^{-\gamma }
\end{equation}

The anomalous cross field diffusion coefficient is 
\begin{equation}
\label{eq:kappaa1}
\kappa_{\rm a}  = \kappa_{\perp} + \frac{2
\,D_B\,\kappa_\|}{\lambda_{\rm L}\,\ln\Lambda}
\end{equation}
for time-scales longer than the particle-field de-correlation time
\glossary{\glo{$\kappa_{\rm a}$}{anomalous cross field diffusion
coefficient (Eqs. \ref{eq:kappaa1}, \ref{eq:kappaLambda})}}
\citep{1995A&A...302L..21D}. $D_B = \delta B^2(l_B)\,
\lambda_\|/(4\,B^2)$ is the field line wandering diffusion
coefficient, and $\lambda_{\rm L} = B^2\,l_\perp^2/(\lambda_\|\,\delta
B^2(l_B))$ is the Liapunov length-scale\footnote{Strictly speaking,
only the perpendicular component of the field fluctuations $\delta
B_\perp^2(l_B)$ enter the equations for $D_B$ and $\lambda_{\rm
L}$. However, the difference between the perpendicular and total
fluctuation levels are not very large. Since also the diffusion
coefficients depend on the perpendicular magnetic fluctuations, one
can interpret $\delta B^2$ as the perpendicular fluctuations right
from the beginning and get a consistent formalism.}. This is the
amplification length of (the expectation value of the square of) a
small displacement of a particle from its initial field line while
travelling along its new field line. The characteristic perpendicular
length-scale $l_\perp$ of the size of {\it chaotic}, non-trivially
twisted magnetic field fluctuations is expected by us to be of the
order of $l_B$, although arguments exist that it might be much
smaller\footnote{Pertubatively (to lowest order in $\delta \vec{B}^2$)
one can show that $l_\perp$ is given by the second Taylor coefficient
of the magnetic field autocorrelation function perpendicular to the
main field direction: $\langle \vec{\delta B}(\vec{x}) \cdot
\vec{\delta B}(\vec{x}+\vec{\xi}_\perp) \rangle = \delta B^2(l_B)\, (1
- \frac{1}{2}\,\xi_\perp^2/l_\perp^2 + O(\xi_\perp^4))$. More strictly
speaking, $l_\perp$ seems to be the Taylor length of the along the
mean field direction integrated and by $\lambda_\|$ normalised
autocorrelation function (of the {\it chaotic} part of the fluctuation
spectrum). However, if the autocorrelation function is a direct
product of a parallel and perpendicular profile, then this is
identical with the simplified definition given here. Otherwise it can
be expected to be close to that. However, it has been pointed out by
\cite{2001ApJ...562L.129N} that if the {\it chaotic} (topologically
non-trivially twisted) magnetic fluctuations extend from the
turbulence energy injection range down to the smallest turbulence
scale, the anomalous diffusion seems to become very efficient. If all
the magnetic power is in such {\it chaotic} modes the effective cross
field diffusion coefficient seems to be close to the order of the
parallel diffusion coefficient. In the case that the magnetic
turbulence in radio cocoons is completely of this {\it chaotic} nature
our results would be changed in a way that particle escape from radio
cocoons is always very fast. Thus the observational tests of particle
escape mentioned in the introduction may confirm or refute such a
picture. In the following it is assumed that the small-scale
turbulence is dominated by travelling Alfv\'enic waves, which do not
disturb the topological properties of the main field. Only on the
largest scales we do expect strong non-trivial topologies, produced by
the turbulence driving forces. Their strength should be limited (and
therefore possibly confined from the smaller-scales) by reconnection
events. Therefore we expect $l_\perp \sim
l_B$.}.\glossary{\glo{$l_\perp$}{$\langle \vec{\delta B}(\vec{x})
\cdot \vec{\delta B}(\vec{x}+\vec{\xi}_\perp) \rangle = \delta
B^2(l_B)\, (1 - \frac{1}{2}\,\xi_\perp^2/l_\perp^2 + O(\xi_\perp^4))$}
if only the {\it chaotic} (topologically non-trivial twisted) field
fluctuations are regarded.} With these definitions one gets
\begin{eqnarray}
\label{eq:kappaLambda}
\kappa_{\rm a} \!\! &=& \kappa_\perp \,\left( 1 +
\frac{\Lambda^2\,\lambda_\perp^2}{\ln\Lambda\,l_\perp^2}\right)\\ 
\!\! &\approx&
\!\! \frac{v\,l_B\,\delta_0\,\lambda_\|^2}{6\,\eps_{\rm
0}\,\ln\Lambda\,l_\perp^2} \, \left( \frac{p\,c}{Z\,e\,B\,l_B}
\right)^{1-\gamma }\\ 
\!\! &\approx& \!\! 1.2\cdot 10^{27}\,\frac{\rm
cm^2}{\rm s} \frac{v\,\delta_0\,\lambda_\|^2}{c\, \eps_0\,l_\perp^2} \left(
\frac{p\,c}{\rm GeV} \right)^{\!\frac{1}{3}} \left(
\frac{l_B}{{\rm kpc}} \right)^{\!\!\frac{2}{3}} \left( \frac{Z\,B}{\mu
{\rm G}}
\right)^{\!\!-\frac{1}{3}}.\!\!\!\!\!\!\!\!\!\!\!\!\!\!\!\!\nonumber
\end{eqnarray}
The ratio of the anomalous to parallel
diffusion coefficients
\begin{equation}
\kappa_{\rm a}/\kappa_\| = \frac{\delta_0^2\,\lambda_\|^2}{2\,\ln\Lambda\,l_\perp^2} \approx
 0.037\, \delta_0^2\,\lambda_\|^2/l_\perp^2
\end{equation}
is nearly independent of the particle momentum, and it is independent
of the fudge factor $\eps_0$. The numerical result is in good
agreement with detailed Monte-Carlo simulation of particle diffusion
in slab-geometry and homogeneous turbulence by 
\cite*{1999ApJ...520..204G}. These authors find a ratio of
$\kappa_{\rm a}/\kappa_\| = 0.02...0.04$ for a turbulence strength 
comparable with $\delta_0 = 1$ (see their Fig. 3) and a scaling which
seems to be only slightly weaker than $\propto \delta_0^2$ (see their
Fig. 4).

\subsection{Enhanced anomalous diffusion\label{sec:EAD}}

Under certain circumstances the anomalous cross field diffusion is
extremely efficient.  The anomalous diffusion coefficient $\kappa_{\rm
a}$ depends quadratically on $\Lambda$ (Eq. \ref{eq:kappaLambda}),
which is mainly the ratio of large-scale to small scale turbulence
(Eq. \ref{eq:Lambda}). $\Lambda$ is therefore independent of the level
of the turbulence, as long as the spectral shape of the turbulence
energy spectrum is not changed. Increasing the turbulence energy
density of a system by a factor $X_{\rm T}$ increases $\kappa_{\rm a}$ also
by $X_{\rm T}$, due to the dependence of $\kappa_\perp \propto X_{\rm T}$ (or
$\kappa_\| \propto X_{\rm T}^{-1}$) on the small scale
turbulence.\glossary{\glo{$X_{\rm T}$}{increment factor of turbulence
energy density}}

If the slope of the turbulence is changed, drastic changes in
$\kappa_{\rm a}$ can result. If e.g. large-scale kinetic energy is
suddenly injected into a system, it may need a period on the order of
the eddy turnover time $\tau_{\rm T} = l_{\rm T}/v_{\rm T}$ before the
turbulent cascade has also raised the small-scale turbulence
level.\glossary{\glo{$\tau_{\rm T}$}{eddy turn-over time}} During this
period the microscopic diffusion coefficients $\kappa_{\|}$ and
$\kappa_{\perp}$ stay unchanged, but $\kappa_{\rm a}$ is increased by
a factor of $X_{\rm T}^2$.

\section{Cosmic ray escape\label{sec:escape}}

\subsection{Boundary layer penetration\label{sec:layer}}

If a boundary layer of insulating magnetic fields (without any flux
leaving the boundary layer regions) separates the interior from the
surrounding of the source region the escape of cosmic ray particles is
strongly inhibited. Although radio polarization observations of radio
cocoons indicate that magnetic fields are well aligned with the cocoon
surface \citep[e.g.][]{1980MNRAS.193..439L, 1981ApJ...248...87L,
1984AJ.....89.1478S}, it is far from obvious if this implies an
insulating boundary layer. The origin of such a hypothetical layer may
be due to mixing of radio plasma with the ambient gas in a thin
surface mixing layer, which is then dynamically decoupled from the
turbulent interior due to its much higher inertia. The expected shear
between the interior and the mixing layer may amplify and align
magnetic fields into such an insulating boundary layer.

As a simplified model we assume the source region to be spherical with
radius $r_{\rm s}$,\glossary{\glo{$r_{\rm s}$}{source radius in
spherical approximation}} the boundary layer to have a thickness
$d_{\rm b}$\glossary{\glo{$d_{\rm b}$}{thickness of insulating
boundary layer $d_{\rm b} = r_{\rm s} - r_{\rm i}$}}. The layer, which
extends from radius $r_{\rm i} = r_{\rm s} - d_{\rm b}$ to $r_{\rm
s}$, is assumed to be filled with magnetic fields which are mainly
tangential to the source surface at $r_{\rm s}$.\glossary{\glo{$r_{\rm
i}$}{inner radius of (spherical) boundary layer}} Therefore the radial
diffusion coefficient is that of the macroscopic cross field diffusion
$\kappa_{\rm a}$. In a quasi-stationary situation (slow particle
escape) the space density of particles is given by
\begin{equation}
g(r,p) = g_0(p)\frac{r^{-1} - r_{\rm s}^{-1}}{r_{\rm i}^{-1} - r_{\rm s}^{-1}}
\end{equation}
within the boundary layer ($r_{\rm i}<r<r_{\rm s}$), where $g_0(p)$ is
the particle spectrum inside the layer ($r<r_{\rm i}$). From this the
number of CRs within the source region is given by
\begin{equation}
N(p) = \int\!\! dV_{\rm s} \,g(r,p) = \frac{2\,\pi}{3}\, g_0(p)\,r_{\rm
s}\,r_{\rm i}\,(r_{\rm s} +r_{\rm i})\,,
\end{equation}
and the CR escape rate through the source surface area $A_{\rm s}$ is given by
\begin{equation}
\dot{N}(p) = A_{\rm s}\,\kappa_{\rm a}\,\nabla g(r,p)|_{r=r_{\rm s}} =
- \frac{4\,\pi\, \kappa_{\rm a} g_0(p)}{r_{\rm i}^{-1} -r_{\rm s}^{-1}}\,.
\end{equation}
The escape frequency follows to be
\begin{eqnarray}
\nu_{\rm layer}(p) &=& - \frac{\dot{N}(p)}{N(p)} = \frac{6\,\kappa_{\rm
a}(p)}{r_{\rm s}^2 - r_{\rm i}^2} = \frac{3\,\kappa_{\rm a}(p)}{r_{\rm
s}\,d_{\rm b}\,(1-d_{\rm b}/(2\,r_{\rm s}))}\\
&\approx& \frac{4.0}{\rm Gyr} \,\frac{v\,\delta_0}{c\,
\eps_0} 
\frac{\lambda_\|^2\,l_B^{\frac{2}{3}}\,{\rm kpc}^{\frac{4}{3}}}{r_{\rm s}\,d_{\rm b}\,l_\perp^2}
\left( \frac{p\,c}{\rm GeV} \right)^{\!\frac{1}{3}}
 \left(
\frac{Z\,B}{\mu {\rm G}}
\right)^{\!\!-\frac{1}{3}}.\!\!\!\!\!\!\!\!\!\!\!\!\!\!\!\!\nonumber
\end{eqnarray}

It should be noted that enhanced anomalous diffusion due to a sudden
raise of the large-scale turbulence level by a factor $X_{\rm T}$
increases the particle escape rate through a boundary layer by a
factor $X_{\rm T}^2$ during the period of not fully developed
turbulence, and $X_{\rm T}$ afterwards (see Sect. \ref{sec:EAD}).

\subsection{Flux tube escape route\label{seq:paralell}}

\begin{figure}[t]
\begin{center}
\psfig{figure=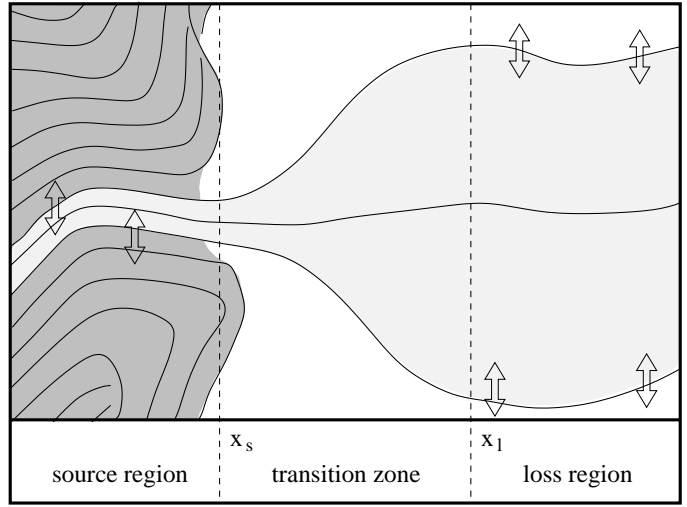,width=\figwidth,angle=0}
\end{center}
\caption[]{\label{fig:geo} Geometry of the inter-phase region. The
left ($x<x_{\rm s}$) and right regions ($x_{\rm l}<x$) are the source
and loss regions of the particle. A bundle of field lines connect
these regions through a transition zone ($x_{\rm s}< x <x_{\rm l}$).}
\end{figure}

If some magnetic flux $\phi_B$ leaks from the source to the loss
region, the CR can escape following the field lines. A possible
situation is sketched in Fig. \ref{fig:geo}. The leakage requires that
the particles of the source region enter the inter-phase flux tube
either by perpendicular diffusion from disconnected regions, or by
travelling along the flux tube from a distant reservoir. Then they
diffuse through a transition zone between the two phases and finally
leave the tube in the loss region by cross field diffusion (or
escape to infinity along the field line).

For simplicity it is assumed that within the source and the loss
regions all parameters of Eq. \ref{eq:difff} are spatially
constant. The source function $q$ is given by $f_{\rm s}$, which is
the average density of particles within the source (per magnetic flux)
divided by the time $\tau_{\rm s}$ a particle needs to cross-field
diffuse over the distance given by the diameter of the flux
tube.\glossary{\glo{$f_{\rm s}$}{average density of particles ($f$)
within the source}} \glossary{\glo{$\tau_{\rm s}$}{flux tube escape
time in source region}} The typical time a particle requires to enter
or leave a flux tube with flux $\phi = \pi\,r^2\,B$ and radius $r
\approx \lambda_\perp/2$ is given by $\tau = r^2/(4\,\kappa_{{\rm a}})
= \phi/(4 \,\pi \, B\, \kappa_{{\rm a}})$.  In the loss region $q=0$.
In the transition zone $\kappa_\|$ and $B$ might be functions of the
position, but no particle losses or sources are assumed
there. Furthermore, any energy loss processes are ignored. These
conditions lead to the following parameters:
\begin{eqnarray}
q(x,p) &=& H(x_{\rm s}-x) \, f_{\rm s}(p)/\tau_{\rm s}(p)\,,\\
\tau(x,p) &=& H(x_{\rm s}-x) \, \tau_{\rm s}(p) + H(x-x_{\rm l})\,
\tau_{\rm l}(p)\,,\\
\dot{p}(x,p) &=& 0\,,\\
B(x) &=&  H(x_{\rm s}-x)\,B_{\rm s} + H(x-x_{\rm l}) \,B_{\rm l}\nonumber\\
&&+ H(x-x_{\rm s})\,H(x_{\rm l}-x)
\,B_{\rm t}(x) \,,\,\mbox{and}\,,\\
\kappa_\|(x,p) &=& H(x_{\rm s}-x)\,\kappa_{\rm s}(p) + H(x- x_{\rm l}) \,\kappa_{\rm l}(p)\nonumber\,,\\
&&+ H(x- x_{\rm s})\,H(x_{\rm l}-x) \,\kappa_{\rm t}(x,p)\,.
\end{eqnarray}
$H(x)$ is the Heaviside step function.  Since the momentum is
conserved ($\dot{p}(x,p) = 0$), Eq. \ref{eq:difff} can be solved
independently for all momenta. The explicit dependency on $p$ is
dropped in the following. The solution of Eq. \ref{eq:difff} with the
parameters given above, smooth field lines ($dB/dx = 0$ at $x_{\rm s}$
and $x_{\rm l}$) and the boundary conditions $f(-\infty) = f_{\rm s}$,
$f(\infty) = 0$ is
\begin{displaymath}
f(x) \!\!= \!\!\left\{\!\!
\begin{array}{ll}
f_{\rm s} - (f_{\rm s} - f_{\rm a})\, \exp ((x-x_{\rm
s})/l_{\rm s}) \!\!& ; x_{\ }\!\!<\!\!x_{\rm s}\\
\left[ B_{\rm s}\, f_{\rm a} - (B_{\rm s}\, f_{\rm a} - B_{\rm l}\, f_{\rm b})\,
\sigma (x) \right]/B_{\rm t}(x) \!\! & ; x_{\rm s}\!\!<\!\!x_{\ }\!\!<\!\!x_{\rm l}\\
f_{\rm b} \, \exp((x_{\rm l}- x)/l_{\rm l}) \!\! & ; x_{\rm l} \!\!<\!\! x
\end{array}\!\!\right.,
\end{displaymath}
\begin{equation}
\label{eq:solution}
\end{equation}
with
\begin{eqnarray}
\label{eq:llls}
l_{\rm s} &=& \sqrt{\kappa_{\rm s}\, \tau_{\rm s}}\,,\,\,\,
%\\
l_{\rm l} = \sqrt{\kappa_{\rm l}\, \tau_{\rm l}}\,,\\
f_{\rm a} &=& f_{\rm s}/ \left[ 1 + 1/\left( \frac{\kappa_{\rm s}\,
I}{l_{\rm s}\, B_{\rm s}} + \frac{\kappa_{\rm s}\,l_{\rm l}\, B_{\rm
l}}{\kappa_{\rm l}\,l_{\rm s}\, B_{\rm s}} \right) \right]\,,\\
f_{\rm b} &=& f_{\rm s}/\left[ \frac{\kappa_{\rm l}\, I}{l_{\rm l}\,
B_{\rm s}} + \frac{B_{\rm l}}{B_{\rm s}} +\frac{\kappa_{\rm l}\,
l_{\rm s}}{\kappa_{\rm s}\, l_{\rm l}} \right]\,,\\
\sigma(x) &=& \frac{1}{I}\, \int_{x_{\rm s}}^{x}\!\!\!\!\! dx'\,
\frac{B_{\rm t}(x')}{\kappa_{\rm t}(x')}\,,\,\mbox{and}\,\,\,\,
%\\
%
I =  \int_{x_{\rm s}}^{x_{\rm l}}\!\!\!\!\! dx'\,
\frac{B_{\rm t}(x')}{\kappa_{\rm t}(x')}\,.
\end{eqnarray}
\glossary{\glo{$l_{\rm s}$ ($l_{\rm l}$)}{characteristic distance a CR travels
before leaving a flux tube in the CR source (loss) region
(Eq. \ref{eq:llls})}} \glossary{\glo{$N_{\rm
s}$}{number of CR in source region}}

\begin{figure}[t]
\begin{center}
\psfig{figure=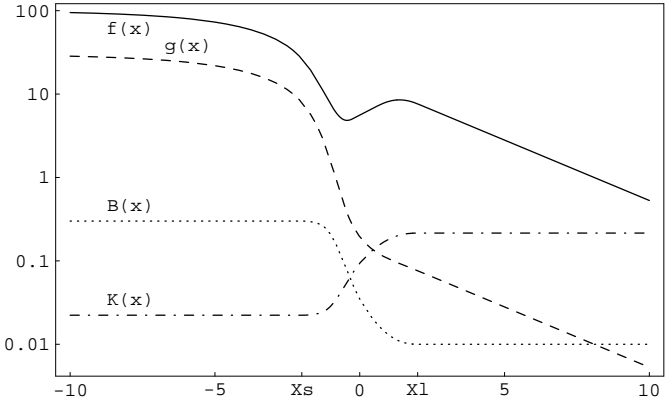,width=\figwidth,angle=0}
\end{center}
\caption[]{Profiles of $f(x)$, $g(x) = B(x) \,f(x)$,
$B(x)$, and $\kappa_{\|}(x)$. The parameters are 
$f_{\rm s} = 100$, $B_{\rm s} = 0.3$, $B_{\rm l} = 0.01$, $x_{\rm s} =
-2$, $x_{\rm l} = 2$, $l_{\rm s} = 3$, $l_{\rm l} = 3$.
The field in the transition zone is parametrised by 
$B_{\rm t}(x) = [B_{\rm l} - (B_{\rm l} - B_{\rm s}) \exp(-
\sqrt{B_{\rm s}/B_{\rm l}} \,(x-x_{\rm s})^2/(x_{\rm l}-x)^2)] + B_{\rm s} - (B_{\rm
s} - B_{\rm l}) \exp(- \sqrt{B_{\rm l}/B_{\rm s}}
\,(x-x_{\rm l})^2/(x_{\rm s}-x)^2)/2$.
Further $\kappa_\|(x) = 0.01\,B^{-2/3}(x)$ is assumed.\label{fig:prof1}}
\end{figure}

\begin{figure}[t]
\begin{center}
\psfig{figure=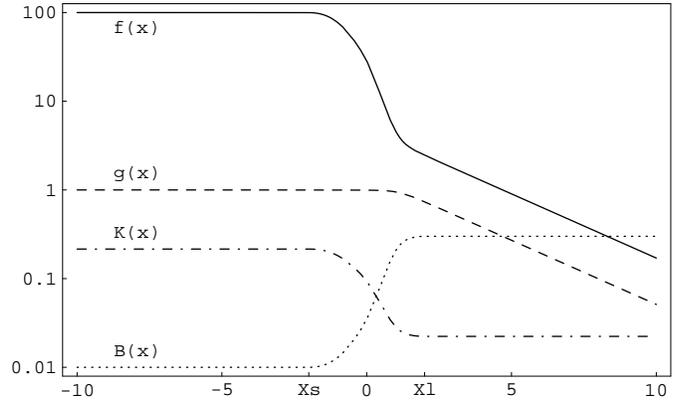,width=\figwidth,angle=0}
\end{center}
\caption[]{\label{fig:prof2} As Fig. \ref{fig:prof1}, but $B_{\rm s} = 0.01$, $B_{\rm l} = 0.3$.}
\end{figure}

An example density profile is shown in Fig. \ref{fig:prof1} for
strong fields in the source and weak field in the loss region, and in
Fig. \ref{fig:prof2} for the reversed geometry.

If one introduces the spatial average within the transition zone by
$\langle A(x) \rangle = \int_{x_{\rm s}}^{x_{\rm l}} dx A(x)/l_{\rm
t}$, where $l_{\rm t} = x_{\rm l}-x_{\rm s}$, one can write
\begin{eqnarray}
I &=&  \bar{B}_{\rm t}\,l_{\rm t} /\bar{\kappa}_{\rm t} = 
 \bar{B}_{\rm t} \,\sqrt{\tau_{\rm t} /\bar{\kappa}_{\rm t}}
 \,\mbox{, where}\\
\bar{B}_{\rm t} &=& \langle B \rangle \,,
\bar\kappa_{\rm t} = \bar{B}_{\rm t} /\langle B/\kappa_{\rm t}
 \rangle \,, \mbox{ and}\;
\tau_{\rm t} = l_{\rm t}^2/\bar\kappa_{\rm t}\,.
\end{eqnarray}
If $B$ changes by a large factor in the transition zone one expects
$\max(B_{\rm s},B_{\rm l}) > \bar{B}_{\rm t} \gg \min(B_{\rm s},B_{\rm
l})$. Due to the expected anti-correlation of $B$ and $\kappa_\|$ also
$\max(\kappa_{\rm s},\kappa_{\rm l}) \gg \bar{\kappa}_{\rm t} >
\min(\kappa_{\rm s},\kappa_{\rm l})$ is likely.

The total particle loss of the source into the loss region is given by
\begin{equation}
\label{eq:loss} 
- \dot{N}_{\rm s\|} \!\!=\!\!\! \int_{x_{\rm l}}^{\infty}\!\!\!\!\!\!\!\! dx\,
  \frac{f(x)}{\tau_{\rm l}} \!\!=\!\! 
 \frac{f_{\rm s}\, \phi_B \, B_s}{B_{\rm
  s}\,\sqrt{\tau_{\rm s}/\kappa_{\rm s}}+ \bar{B}_{\rm t}
  \, \sqrt{\tau_{\rm t}/\bar\kappa_{\rm t}}+ B_{\rm l}
  \,\sqrt{\tau_{\rm l}/\kappa_{\rm l}}}
\end{equation}
If several loss channels exist with similar properties ($\tau_{\rm
s}$, $\tau_{\rm l}$) then $\phi_B$ in Eq. \ref{eq:loss} can be
replaced by the sum of the absolute fluxes of these channels. In this
case the flux leaving the source region is $\phi_{\rm B} =
A_s\,\eta_{\rm s}\,B_{\rm s}$, where $\eta_{\rm s}$ is the fraction of
the source surface ($A_{\rm s}$) penetrated by
inter-phase magnetic flux.  \glossary{\glo{$A_{\rm s}$}{source region
surface area}} \glossary{\glo{$\eta_{\rm s}$}{fraction of $A_{\rm s}$
occupied by inter-phase magnetic flux}} After defining the
characteristic length-scale of the source by $L_{\rm s} = V_{\rm
s}/A_{\rm s}$, where $V_{\rm s}$ is the source volume,
\glossary{\glo{$V_{\rm s}$}{source volume}} \glossary{\glo{$L_{\rm s}=
V_{\rm s}/A_{\rm s}$}{characteristic length of source}} and noticing
that $N_{\rm s} = V_{\rm s}\,g_{\rm s} = V_{\rm s}\,f_{\rm s} \,B_{\rm
s}$ one can write the typical particle escape frequency $\nu_{\|}$ as
\begin{equation}
\label{eq:escapetime}
\nu_{\|} = -\frac{\dot{N}_{\rm s\|}}{N_{\rm s}} = 
 \frac{\eta_{\rm s} \, B_s/ L_{\rm s}}{ B_{\rm
  s}\,\sqrt{\tau_{\rm s}/\kappa_{\rm s}}+ \bar{B}_{\rm t}
  \, \sqrt{\tau_{\rm t}/\bar\kappa_{\rm t}}+ B_{\rm l}
  \,\sqrt{\tau_{\rm l}/\kappa_{\rm l}} }\,.
\end{equation}
\glossary{\glo{$\nu_{\|}$}{CR escape frequency along flux tubes
(Eqs. \ref{eq:escapetime}, \ref{eq:moreloss.2})}}

This formula is simplified if one of the terms in the denominator
dominates. Two cases correspond to typical astrophysical situations:\\ 
\noindent
{Strongly magnetised source region:}
\begin{equation}
\label{eq:strong}
\nu_{\|} \approx \frac{\eta_{\rm s}}{ L_{\rm s}}
 \,\sqrt{\frac{\kappa_{\rm s}}{\tau_{\rm s}}} = 
 \frac{\eta_{\rm s}\,\kappa_{\rm s}}{ L_{\rm s}\,l_{\rm s}}= 
 \frac{\phi_B\,\kappa_{\rm s}}{B_{\rm s}\,  V_{\rm s}\,l_{\rm s}}
\end{equation}
\noindent
{Strongly magnetised loss region:}
\begin{equation}
\label{eq:weak}
\nu_{\|} \approx \frac{\eta_{\rm s}\,B_{\rm s}}{L_{\rm s}\,B_{\rm l} }
 \,\sqrt{\frac{\kappa_{\rm l}}{\tau_{\rm l}}} = 
 \frac{\eta_{\rm s}\,B_{\rm s}\,\kappa_{\rm l}}{ L_{\rm s}\,l_{\rm l}\,B_{\rm l}}= 
 \frac{\phi_B\,\kappa_{\rm l}}{B_{\rm l}\,  V_{\rm s}\,l_{\rm l}}
\end{equation}

If two regions 1 and 2 with volumes $V_1$ and $V_2$ are connected by
some magnetic flux $\phi_{\rm B}$, then the ratio of the particle
exchange frequencies is equal to the inverse volume ratio, as it is
required by detailed balance:
\begin{equation}
\label{eq:det.bal}
\frac{\nu_{\rm 1 \rightarrow 2}}{\nu_{\rm 2 \rightarrow 1}} =
\frac{\eta_{\rm 1} \, B_{\rm 1}/ L_{\rm 1}}{\eta_{\rm 2} \, B_2/ L_{\rm 2}}
= \frac{V_{\rm 2}}{V_{\rm 1}}\,.
\end{equation}

In the following it is assumed that the source field region has the
strongest magnetic fields and therefore Eq. \ref{eq:strong} is used.
If a number of $M_\|$ flux-tubes with diameter $\lambda_\perp$ each
leave the source region the total magnetic flux leaving the source is
$\phi_B = M_\|\, \pi\,B\,\lambda_\perp^2/4$.  Applying the identity
$M_\| = {4\,\eta_{\rm s}\,A_{\rm s}}/({\pi \, \lambda_\perp^2})$ one
finds that the particle loss frequency is then given by
\glossary{\glo{$M_\|$}{number of inter phase flux tubes}} 
\begin{eqnarray}
\label{eq:moreloss.2}
\nu_{\|} &\approx& \frac{4 \, \eta_{\rm s}\, \,\sqrt{\kappa_{\rm a}\,
 \kappa_\|}}{\lambda_\perp\, L_{\rm s}} \\
\label{eq:tmp1.2}
&=& \frac{4 \, \eta_{\rm s}\,l_{\rm B}\,\lambda_\|\, v}{3 \,
\sqrt{2 \, \ln\Lambda} \,\eps_0\, l_\perp\,\lambda_\perp\,
L_{\rm s}}\,
\left(\frac{p\,c}{Z\,e\,B\,l_B} \right)^{1-\gamma }\\ 
&\approx& \frac{0.081}{\rm Myr}\,\eta_{\rm s}\,\frac{v}{c\,\eps_0} \frac{l_B^{2/3}
\,\lambda_\|\,/(\lambda_\perp \,l_\perp \,L_{\rm s})}{{\rm kpc}^{-4/3}} \,
\left(\frac{p\,c/(Z \,B)}{{\rm GeV}/\mu{\rm G}} \right)^{\!\!\frac{1}{3}}\,.
\nonumber
\end{eqnarray}

It should be noted that enhanced anomalous diffusion due to a sudden
raise of the large-scale turbulence level by a factor $X_{\rm T}$ increases
the flux tube particle escape rate by a factor $X_{\rm T}$ only during the
period of not fully developed turbulence, but not afterwards (see
Sect. \ref{sec:EAD}).

If the CRs escape rapidly from the source region, they can excite
plasma waves which again scatter the CRs. This limits the escape
velocity to approximately the Alfv\'en- or thermal velocity, or
whichever is larger \citep{1969ApJ...158..959T,
1979ApJ...228..576H}. For the high energy part of the CR spectrum the
limiting velocity can be even larger, since the small number density
of high energy CRs in typical power-law distributions does not lead to
an efficient excitation of the scattering waves
\citep[e.g.][]{2001ApJ...553..198F}. However, an important quantity is
the CR escape velocity at the surface of the source region:
\begin{eqnarray}
\label{eq:vcr}
v_{\rm CR} &=& \frac{\kappa_{\rm s}}{l_{\rm s}} =
\frac{4\,v}{3\,\eps_0\,\sqrt{2\ln\Lambda}}\, \frac{l_B\,
\lambda_\|}{\lambda_\perp\,l_\perp}\,
\left(\frac{p\,c}{Z\,e\,B\,l_B}\right)^{1-\gamma}\\
&\approx & 80 \,\frac{\rm km}{\rm s}\, 
\frac{v}{\eps_0\,c} \,
\frac{l_B^{5/3} \,\lambda_\|}{\lambda_\perp\, l_\perp\,{\rm
kpc}^{2/3}}\,
\left(\frac{p\,c/(Z \,B)}{{\rm GeV}/\mu{\rm G}} \right)^{\!\!\frac{1}{3}}.
\nonumber
\end{eqnarray}
As long as this velocity is below the limiting speed, which for radio
galaxy cocoons should be the environmental sound speed, CR
self-confinement due to wave excitation should be negligible.

\subsection{Cross field escape route\label{seq:perpendicular}}

\begin{figure}[t]
\begin{center}
\psfig{figure=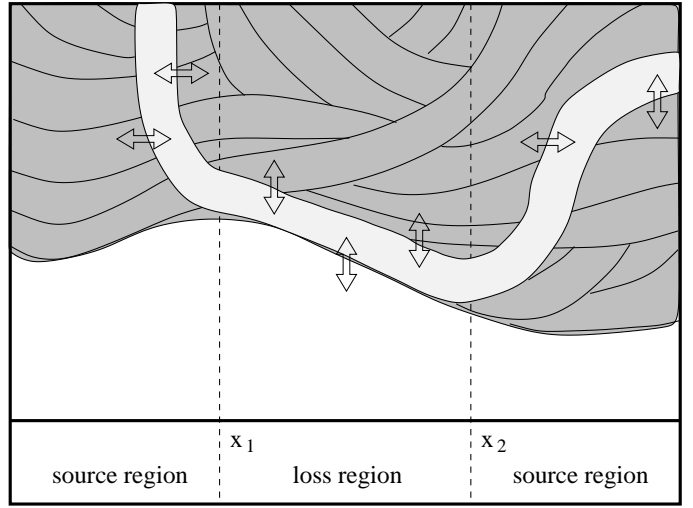,width=\figwidth,angle=0}
\end{center}
\caption[]{\label{fig:geoperp} Geometry of a flux tube which is part of
the source boundary. The
left ($x<x_{\rm 1}$) and right regions ($x_{\rm 2}<x$) are the source
and loss regions of the particles. In between, escape into the other
phase is possible.}
\end{figure}

\begin{figure}[t]
\begin{center}
\psfig{figure=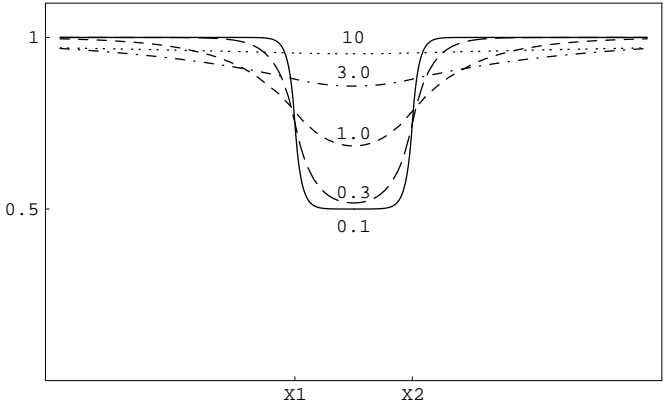,width=\figwidth,angle=0}
\end{center}
\caption[]{Profile of $f(x)$ for $\sigma_\perp = 0.5$, $\lambda_\| =2$ and different $l_s$
as labelled in the graph.\label{fig:fperp}}
\end{figure}

In order to model the cross field escape frequency an individual flux tube
of diameter $\lambda_\perp$ is investigate first, which touches the
inter phase surface on a length-scale $\lambda_\|$ (see
Fig. \ref{fig:geoperp}). The particle distribution function along this
tube is governed by Eq. \ref{eq:difff} if one adopts the following
parameters:
\begin{eqnarray}
q(x,p) &=& (1 - \sigma_\perp(x)) \, f_{\rm s}(p)/\tau_{\rm s}(p)\,,\\
\tau(x,p) &=& \tau_{\rm s}(p)\,,\\
\dot{p}(x,p) &=& 0\,,\\
B(x) &=&  B_{\rm s}\\
\kappa_\|(x,p) &=& \kappa_{\rm s}(p) \nonumber\,.
\end{eqnarray}
$\sigma_\perp(x)$ is the fraction of the flux tubes surface which is
in contact with the other phase. $\sigma_\perp(x) = 0$ for $x < x_1$
or $x>x_2$. Again, the notation of the momentum dependence is dropped for
convenience. The solution can be found with the help of the method of
Green's functions and is given by
\begin{equation}
f(x) = f_{\rm s} \,\left[ 1 - \frac{1}{2\, l_s}
\int_{x_1}^{x_2}\!\!\!\!  dx' \,
\sigma_\perp (x') \, \exp(-| x-x'|/l_s)\right]
\end{equation}
In the following it is assumed that $\sigma_\perp(x) = {\rm const}$ within
the length $\lambda_\| = x_2 -
x_1$. \glossary{\glo{$\sigma_\perp$}{fraction of flux tube surface in
contact with loss region}} In this case
\begin{eqnarray}
f(x) &=& f_{\rm s} \left[ 1- \frac{\sigma_\perp}{2}\, \left( h\left(\frac{x-x_1}{l_{\rm s}}
\right) + h\left(\frac{x_2-x}{l_{\rm s}} \right) \right) \right],\,\mbox{with}\nonumber\\ 
h(z) &=& {\rm sgn} (z) \,(1 - e^{-|z|})\,,
\end{eqnarray}
which is shown in Fig. \ref{fig:fperp}.  The particle escape rate
(into the other phase) of this filament is then
\begin{equation}
 - \dot{N}_\perp = \frac{f_{\rm s} \, \lambda_\|\, \sigma_\perp}{\tau_{\rm s}}
\left[ 1 - \frac{\sigma_\perp}{y} \,(y +  e^{- y} -1)  \right]\,,
\end{equation}
where \glossary{\glo{$y = \lambda_\|/{l_{\rm s}}$}{field correlation
length over particle diffusion length (Eq. \ref{eq:def_y})}}
\begin{equation}
\label{eq:def_y}
y = \frac{\lambda_\|}{l_{\rm s}} = \frac{\lambda_\|}{\sqrt{\kappa_{\rm
s}\, \tau_{\rm s}}} = \frac{\lambda_\|^2}{l_\perp\,\lambda_\perp} \,
\frac{4 \, \delta_0}{\sqrt{2\, \ln \Lambda}}.
\end{equation}
Usually $y\ll 1$, otherwise the situation resembles more the
isolating boundary layer case discussed in Sect. \ref{sec:layer}.
The number $M_\perp$ \glossary{\glo{$M_\perp$}{number of flux tubes
forming the boundary}} of flux tubes which touch that part of the
surface of the source region, which is not intersected by inter phase
magnetic flux, is
\begin{equation}
M_\perp =  \frac{A_{\rm s}\, (1 -\eta_{\rm s}) }{\pi\, \lambda_\perp \,
\lambda_\| \, \sigma_\perp}\,.
\end{equation}
This leads to perpendicular escape frequency of
\glossary{\glo{$\nu_\perp$}{perpendicular escape frequency}}
\begin{equation}
\label{eq:nuperp}
\nu_\perp = - \frac{M_\perp \, \dot{N}_\perp}{N_{\rm s}} = \nu_{\perp,
0} \,(1 -\eta_{\rm s}) \left[ 1- \frac{\sigma_\perp}{y} \,(y +  e^{- y} -1) \right]\!,
\end{equation}
with
\begin{eqnarray}
\label{eq:nuperp0.1}
\nu_{\perp, 0} &=& \frac{A_{\rm s} \, \lambda_\perp}{4\, V_{\rm s}\,
\tau_{\rm s}} = \frac{4\,\kappa_{\rm a}}{L_{\rm s}\, \lambda_\perp}\\
\label{eq:nuperp0.2}
&=& \frac{2\,\delta_0}{3\, \eps_0\, \ln \Lambda} \, \frac{l_B\,
\lambda_\|^2\, v}{l_\perp^2\, \lambda_\perp \, L_{\rm s}}\,
\left(\frac{p\, c}{Z\, e\, B\, l_B} \right)^{\!\!\frac{1}{3}}\\
&\approx& \frac{0.016}{{\rm Myr}} \, \frac{\delta_0\, v}{\eps_0\, c} \,
\frac{l_B^{2/3}\, \lambda_\|^2/(l_\perp^2 \, \lambda_\perp\,L_{\rm
s})}{{\rm kpc}^{-4/3}} \left( \frac{p\,c / (Z\,B)}{{\rm
GeV}/\mu{\rm G}}
\right)^{\!\!\frac{1}{3}}\!\!\!\!\!\!\!\!\!\!\!\!\!\!\!\!\!\!\!\!\!\!\!\!\!\nonumber
\end{eqnarray}
The reverse process, in which particles enter the strongly magnetised
region, has a frequency which can be calculated from the escape
frequency (Eq. \ref{eq:nuperp}) and the condition of detailed balance
(Eq. \ref{eq:det.bal}). 

It should be noted that enhanced anomalous diffusion due to a sudden raise of the
large-scale turbulence level by a factor $X_{\rm T}$ increases the cross
field particle escape rate by a factor $X_{\rm T}^2$ during the period of
not fully developed turbulence, and $X_{\rm T}$ afterwards (see
Sect. \ref{sec:EAD}).

\subsection{The best escape route}

The most efficient escape route can be found by comparing the escape
frequencies for the flux tube and the cross field escape route:
\begin{eqnarray}
\label{eq:nunu1}
\frac{\nu_\perp}{\nu_\|} &=& \sqrt{\frac{\kappa_{\rm
a}}{\kappa_{\|}}} \frac{1-\eta_{\rm s}}{\eta_{\rm s}} \, \left[ 1 -
\frac{\sigma_\perp}{y} \,(y +  e^{- y} -1) \right]\\ 
\label{eq:nunu3}
&\approx& \frac{\delta_0\, \lambda_\|}{5\, l_\perp\, \eta_{\rm s}}
\end{eqnarray}

During the injection phase of turbulence (a few $\tau_{\rm T}$) the ratio
$\nu_\perp/\nu_\|$ is increased by an additional factor of $X_{\rm T}$ over
the estimates given in Eq. \ref{eq:nunu3}. Under
certain conditions ($\delta_0 < 5\, \eta_{\rm s} <  X_{\rm T}\, \delta_0$)
the dominant escape mechanism can switch temporarily from flux tube
escape to cross field escape during the onset of turbulence.

\section{CR escape from radio plasma \label{sec:appl}}

\subsection{A quiet environment}
The old radio plasma cocoon of a small radio galaxy in the central IGM of
a galaxy cluster might be of approximately spherical shape. Here the
following parameters are assumed: $B = 10\, \mu$G, $l_B = \lambda_\| =
3\,$kpc, $l_\perp = \lambda_\perp = 1\,$kpc, and a cocoon diameter of 50 kpc.
The turbulence inside old radio plasma is likely the turbulence
induced by the environment, and therefore assumed to be
Kolmogorov-like. The diffusion coefficients of 10 GeV CR particles are
then $\kappa_\| = 7\cdot 10^{28}\,{\rm cm^2\,
s^{-1}}/(\eps_0\,\delta_0)$, $\kappa_\perp = 1.7\cdot 10^{16}\,{\rm
cm^2\, s^{-1}}\,\eps_0\,\delta_0$, and $\kappa_{\rm a} = 1.9\cdot
10^{28}\,{\rm cm^2\, s^{-1}}\,\delta_0/\eps_0$.  If the radio cocoon
has a boundary layer of thickness $d_{\rm b} = 5$ kpc, which is
isolating the interior from the exterior due to the lack of any
magnetic flux exchange, then the escape frequency of these CRs is
$\nu_{\rm layer} = 1.7\,{\rm Gyr}^{-1}\,\delta_0/\eps_0$. This is
extremely slow for a low level of magnetic turbulence ($\delta_0 <
0.01$).

If such a isolating layer does not exist, CR particles can 
reach the surface much easier and leave by cross field diffusion (over a short
distance) or by following open field lines. Since only reconnection
events of the fossil fields with the much weaker environmental fields
could have opened the magnetic topology, $\eta_{\rm s} \ll 1$ is
assumed due to the possible rarity of such events. Under such
conditions the escape frequencies of 10 GeV particles are $\nu_{\|}
\approx 57\,{\rm Gyr}^{-1}\, \eta_s/\eps_0$, and $\nu_{\perp} \approx
31\,{\rm Gyr}^{-1}\, \delta_0/\eps_0$. The CR streaming velocity
along open flux tubes is 460 km/s$\, /\eps_0$ and therefore likely
much below the cluster sound speed, so that CR self-confinement is not
important for these particles.  For a low level of magnetic turbulence
($\delta_0 < 0.01$) and a mostly closed field topology ($\eta_{\rm s}
< 0.01$) the escape of 10 GeV particles requires several Gyr.  If one
considers further that such radio plasma is buoyant and therefore
leaves the cluster centre within a few 100 Myr
\citep{2001ApJ...554..261C, 2002A&A...384L..27E}, the fraction of CRs
injected into a cluster core is expected to be small under such
circumstances. This can impose constraints on several of the
theoretical considerations listed in Sect. \ref{sec:motiv}, which rely
on an efficient escape of CR electrons, positrons or protons from
radio cocoons into the centre of galaxy clusters.

Further, during the buoyant rise of a radio cocoon through a
cluster atmosphere, the cocoon expands adiabatically. If all magnetic
length-scales scale linearly with the cocoon size $L_{\rm s}$, and the
magnetic field strength decreases adiabatically as $B \propto L_{\rm
s}^{-2}$ then all escape frequencies discussed here ($\nu_{\rm
layer}$, $\nu_{\|}$, $\nu_{\perp}$) decrease according to $\nu \propto
L_{\rm s}^{-2/3}$. This implies that CRs are better confined in a
larger cocoon, or that CR escape is most rapidly at an early stage of
the cocoon's buoyant voyage through the intra-cluster medium.

\subsection{Suddenly injected turbulence} If a cluster merger
event suddenly injects large-scale turbulence the radio plasma can
become transparent even for low energy CR particles.  The merger might
produce turbulent flows with velocities of $v_{\rm T} \approx
1000\,$km/s on a scale of $l_{\rm T} \approx 100\,$kpc which increase
the turbulent magnetic energy density on large-scales by a factor of
$X_{\rm T} = 10$ from initially $\delta_0 = 0.01$ for example. For
roughly an eddy turnover time $\tau_{\rm T} = l_{\rm T} / v_{\rm T}
\approx 100\,$ Myr the small scale turbulence is not increased in
regions far from shock waves.  During only a tenth of this period (10
Myr, which is of the order the turbulent cascades needs to transfer
energy from the scales of the radio cocoon (10 kpc) down to the CR
gyro-radius length scales) an enhanced anomalous cross field escape
should allow roughly 30 \%$/ \eps_0$ of the 10 GeV CR particles
initially confined in the radio cocoon to escape. A similar number of
the CRs would escape in the remaining 90 Myr of fully developed
turbulence.  Significant losses would occur on even shorter timescales
in the case that the pitch angle scattering efficiency is low ($\eps_0
\ll 1$). The total loss of particles escaping along inter phase flux
tubes is roughly 6 \%$\, (\eta_{\rm s}/0.01) / \eps_0 \ll 1$ during
$0.1\,\tau_{\rm T}$, and likely negligible if only a few kpc$^2$ of
the cocoons surface is opened ($\eta_{\rm s}\ll 0.01$). If the cocoon
would have an isolating boundary layer of thickness 5 kpc, then only
1.7 \% of the CR particles would be released during $0.1\,\tau_{\rm
T}$.

\subsection{Merger shock waves}

Cluster mergers happen frequently \citep{1995ApJ...447....8M,
1999ApJ...511...65J, 2001A&A...378..408S} so that shock waves appear
often in clusters \citep{1998ApJ...502..518Q, 2000ApJ...542..608M},
which should raise the turbulence level on all length scales during
their passage. Therefore the strongly enhanced anomalous diffusivity
discussed above is not expected to appear in this case. However, the
turbulence level should increase substantially, leading to some
enhancement of the particle escape rate, and the typical lengthscale
$L_{\rm s}$ the particles have to travel before leaving the source
also decreases substantially. Both effects in combination should lead
to efficient particle escape.

The escape length $L_{\rm s} = V_{\rm s}/A_{\rm s}$ decreases due to
two effects: First, the radio plasma volume shrinks by a factor
$C=V_2/V_1 = (P_1/P_2)^{3/4}$ \glossary{\glo{$C$}{volume shrinking
factor of a shock compressed radio cocoon}} for radio plasma with a
relativistic equation of state which is adiabatically (due to the high
internal sound speed) compressed in an environmental shock wave
with pressure jump $P_2/P_1$. Second, the shock wave should disrupt
the radio plasma into filamentary or torus-like morphologies, as seen
in numerical simulations \citep{2002MNRAS.331.1011E} and in high
resolution radio maps of some cluster radio relics
\citep{2001AJ....122.1172S}. The numerical simulations show that a
spherical radio cocoon transforms into a torus with major radius equal
to the radius of the original sphere (the torus diameter is roughly
that of the former sphere). Using the dimensions of the torus
given by this observation and the above given volume shrinking factor
$C$ one finds that the characteristic length-scale $L_{\rm s} = V_{\rm
s}/A_{\rm s}$ decreases due to shock compression and change of the
morphology by a factor
\begin{equation}
\frac{L_{\rm s,2}}{L_{\rm s,1}} = \sqrt{\frac{3\,C}{4\,\pi}} \,\left[
1 + \sqrt{\frac{4\,C}{3\,\pi^2}} \right]^{-1} \approx \frac{1}{2} \,\sqrt{C}\,.
\end{equation}
Thus, for a typical merger shock wave with Mach number of the order 3,
and pressure jump roughly 10 the characteristic length scale decreases
by a factor of nearly 6. If at the same time the level of turbulence
increases by an order of magnitude, the cross field diffusion escape
rate is enlarged by nearly two orders of magnitude. Therefore also for
radio plasma passing through a shock wave efficient CR particle escape
is expected.

\section{Discussion\label{sec:diss}}
A model has been presented describing the escape of CR particles from
regions which are poorly magnetically connected to the
environment. The main application of this theoryis to cocoons of radio
galaxies.

In order to apply this model quantitatively to astrophysical systems,
several poorly known parameters have to be determined, like the
geometry of the source ($V_{\rm s}$, $A_{\rm s}$, $L_{\rm s}$,
$\eta_{\rm s}$, $\sigma_{\rm s}$), the field fluctuation length-scales
($\lambda_\|$, $\lambda_\perp$, $l_\perp$, $l_B$), the nature and the
level of the turbulence ($\delta_0$), and the coupling strength
between small-scale magnetic turbulence and the gyro-motion of charged
particles ($\eps_0$). Several of these parameters are closely related
and can roughly be estimated (e.g. the characteristic turbulence
length-scales). Other parameters will remain unknown until a truly
detailed knowledge of radio plasma is available.

However, even without detailed knowledge of these parameters,
insight into the qualitative behaviour of CR transport between
different phases is provided:
\begin{itemize}
\item An isolating magnetic layer at the surface of the CR confining
region can efficiently suppress CR escape compared to the case in which
flux tubes from the source interior touch (or even leave) the source surface.
\item CR escape from a mostly magnetically confined region 
always requires parallel and cross field particle transport.
\item The escape frequency of cosmic rays travelling along interphase
flux tubes is independent of the level of turbulence
(Eqs. \ref{eq:moreloss.2}-\ref{eq:tmp1.2}).
\item The cross field escape frequency increases linearly with increasing
turbulence level (Eqs. \ref{eq:nuperp}-\ref{eq:nuperp0.2}).
\item The parallel and perpendicular
escape frequencies are mainly determined by the properties of the strong field region.
\item All escape frequencies considered here have an identical scaling
with the CR momentum, for any turbulence spectra. E.g. for Kolmogorov
turbulence it is $\nu_{\|/\perp} \propto p^{1/3}$, and for Kraichnan
turbulence it is $\nu_{\|/\perp} \propto p^{1/2}$.
\item In the case where the pitch angle scattering frequency is
strongly reduced \citep{2000PhRvL..85.4656C} due to the possible
anisotropic nature of weak magneto-hydrodynamic turbulence
\citep{1994ApJ...432..612S,1997ApJ...485..680G}, this model can still
be applicable as long as the particle transport is diffusive. In such
a case one expects $\eps_0\ll 1$.
\item If the {\it chaotic} (topologically non-trivially twisted) part
of the magnetic fluctuations extends down in the power-spectrum to the
smallest length-scales, the anomalous cross field diffusion may be
much larger than estimated here \citep{2001ApJ...562L.129N} leading to
a very rapid CR escape from radio cocoons. If this is indeed the case
is not yet clear, but may be testable by observations of escaped
particles\footnote{E.g. annihilation of escaped positrons, decaying
pions produced by relativistic protons, or even simply a low level
diffuse synchrotron halo around a bright radio cocoon due to escaping
electrons.}.
\item In contrast to a statement by \cite{2000ApJ...529..513C} it
is shown here that the macroscopic diffusive transport of particles
along a flux bundle with small scale inhomogeneities in the field
strength is reduced by the presence of strong field bottlenecks
(Appendix A).
\item A sudden large-scale turbulence injection can lead temporarily to
the enhanced anomalous diffusion regime. This increases both escape
frequencies, but the cross field escape frequency by a larger factor.
\item Cluster merger events can lead to strong particle losses from
radio plasma. Shock waves disrupt the relativistic plasma into smaller
pieces from which CRs can escape more easily. It should be noted
that cluster merger waves and turbulence themself are able to
accelerate CRs, which can mask (and/or energise) the CRs escaping from
radio plasma.
\item Merger can also lead to efficient particle escape from radio
plasma not directly affected by shock waves. The enhanced anomalous diffusion
due to the large-scale turbulence injected by the merger could also
lead to efficient particle escape. 
\end{itemize}

Given the large astrophysical importance of the problem of cosmic ray
escape from radio plasma cocoons, further studies of this problem
-- theoretically and observationally -- would be of great value.  The
presented work provides a starting point for such investigations.

\acknowledgements I thank Benjamin Chandran, Eugene Churazov,
Sebastian Heinz, Philipp P. Kronberg, and Kandu Subramanian for
discussions and comments on the manuscript. This work was done in the
framework of the EC Research and Training Network {\it The Physics of
the Intergalactic Medium}.

\appendix
\section{Homogenisation of the diffusion equation\label{sec:homo}}
It is instructive to derive the macroscopic transport equation in the
case that the spatial fluctuations in the parameters of
Eq. \ref{eq:difff} can be described by a large-scale $x'$ and a much
smaller scale $\xi$, so that $x = x' + \xi/\epsilon$ and $\epsilon$ is
the ratio of the small to the large-scale. A quantity $A(x) =
A(x',\xi)$ is averaged over the smaller scale by $\langle A \rangle
(x') = \lim_{\xi \rightarrow \infty} \, \,\int_{0}^{\xi} d\xi' \,
A(x',\xi')/\xi$. 
The evolution of the distribution function can be expressed as an
asymptotic series in the small parameter $\epsilon$. To lowest order
in $\epsilon$ this method of homogenisation gives
\begin{equation}
\label{eq:homo}
\frac{\partial \bar{f}}{\partial t} + \frac{\partial }{\partial p}
(\bar{\dot{p}}\,\,\bar{f}) = \frac{\partial }{\partial x} \left(
\bar{\kappa}_\| \frac{\partial (\bar{B}\,\bar{f})}{\bar{B}\, \partial
x} \right) - \frac{\bar{f}}{\bar{\tau}} + \bar{q}\,,
\end{equation}
where the prime was dropped ($x' \rightarrow x$). The details of this kind of
calculation can be found in \cite*{1995QA871.H74......}.
The homogenised parameters are given by
\begin{eqnarray}
\bar{f} &=& \langle f \rangle \\ 
\bar{B} &=& \langle 1/B \rangle^{-1} \\
\bar{\dot{p}} &=& \bar{B}\, \langle \dot{p}/B\rangle \\
\bar{\kappa}_\| &=& \bar{B}/\langle B/\kappa_\| \rangle \label{eq:kappabar}\\
\bar{\tau} &=& \bar{B}^{-1}/ \langle 1/(\tau\,B) \rangle\\ 
\bar{q} &=& \langle q \rangle \,.
\end{eqnarray}
In the case of vanishing small-scale fluctuations of all coefficients,
their homogenised averages are, of course, unchanged by the
homogenisation and Eq. \ref{eq:homo} is identical to
Eq. \ref{eq:difff}.
In the case of small scale variations, the homogenised loss
coefficients $\bar{\dot{p}}$ and $\bar{\tau}$ are weighted towards the
weak field regions, where the flux tube has a larger diameter and
therefore most of the particles reside.
The effective large-scale diffusion coefficient $\bar{\kappa}_\|$ is
reduced compared to the typical small scale coefficient
$\langle{\kappa}_\|\rangle$, due to the harmonic weighting in
Eq. \ref{eq:kappabar}. The weight of these regions is further
increased due to the expected anti-correlation of field strength and
diffusion coefficient.  However, even in the case that $\kappa_\|$
does not change on the micro-scale, its global value is changed due to
field strength fluctuations on the micro-scale: $\bar{\kappa}_\| =
\kappa_\| /(\langle 1/B \rangle \langle B \rangle ) \le \kappa_\|$
(equality only if $B(x,\xi) = B(x)$). If e.g. 10 \% of the micro-scale
has a 100 times enhanced field strength, the averaged diffusion
coefficient is a factor of 10 smaller than the microscopic one, in the
case of a spatially constant $\kappa_\|$, otherwise much more.  This
demonstrates that magnetic traps modify the particle transport even in
the diffusive regime, in contrast to an opposite statement by
\cite{2000ApJ...529..513C}.

\def\glossaryentry#1#2{#1}
\section{Glossary\label{sec:glaossary}}
%\begin{tabular}{l}
\begin{list}{}{}
\glossaryentry{\glo{$A_{\rm s}$}{source region surface area}}{7}
\glossaryentry{\glo{CMB}{Cosmic Microwave Background}}{1}
\glossaryentry{\glo{CR}{cosmic ray}}{1}
\glossaryentry{\glo{$C$}{volume shrinking factor of a shock compressed radio cocoon}}{10}
\glossaryentry{\glo{$\delta_0$}{relative strength of magnetic turbulence at reference scale (Eq. \ref{eq:magturb})}}{4}
\glossaryentry{\glo{$D_{\mu\mu}$}{Fokker-Planck pitch angle diffusion coefficient}}{3}
\glossaryentry{\glo{$d_{\rm b}$}{thickness of insulating boundary layer $d_{\rm b} = r_{\rm s} - r_{\rm i}$}}{6}
\glossaryentry{\glo{$\eps_0$}{fudge factor relating turbulence energy density to pitch angle scattering efficiency (Eq. \ref{eq:epsfunc})}}{4}
\glossaryentry{\glo{$\eps$}{ratio of scattering frequency to gyro-frequency (Eq. \ref{eq:epsfunc})}}{4}
\glossaryentry{\glo{$\eta_{\rm s}$}{fraction of $A_{\rm s}$ occupied by inter-phase magnetic flux}}{7}
\glossaryentry{\glo{$f$}{isotropic particle phase-space density per magnetic flux (Eq. \ref{eq:fdef})}}{3}
\glossaryentry{\glo{$f_{\rm s}$}{average density of particles ($f$) within the source}}{7}
\glossaryentry{\glo{$\gamma$}{spectral index of magnetic turbulence spectrum}}{4}
\glossaryentry{\glo{$g$}{isotropic particle phase space density per volume}}{3}
\glossaryentry{\glo{$\kappa_\|$}{parallel diffusion coefficient (Eqs. \ref{eq:diffpar1}, \ref{eq:diffpar2}, \ref{eq:diffpar3})}}{3}
\glossaryentry{\glo{$\kappa_\perp$}{perpendicular diffusion coefficient (Eqs. \ref{eq:kappaperp1}, \ref{eq:kappaperp2})}}{4}
\glossaryentry{\glo{$\kappa_{\rm a}$}{anomalous cross field diffusion coefficient (Eqs. \ref{eq:kappaa1}, \ref{eq:kappaLambda})}}{5}
\glossaryentry{\glo{$\Lambda$}{Eqs. \ref{eq:Lambda0}, \ref{eq:Lambda}}}{5}
\glossaryentry{\glo{$\lambda_\|$}{parallel correlation length of the field fluctuations}}{5}
\glossaryentry{\glo{$\lambda_\perp$}{perpendicular correlation length of the field fluctuations, typical flux tube diameter}}{5}
\glossaryentry{\glo{$l_B$}{magnetic turbulence reference scale (Eq. \ref{eq:magturb})}}{4}
\glossaryentry{\glo{$l_\perp$}{$\langle \vec{\delta B}(\vec{x}) \cdot \vec{\delta B}(\vec{x}+\vec{\xi}_\perp) \rangle = \delta B^2(l_B)\, (1 - \frac{1}{2}\,\xi_\perp^2/l_\perp^2 + O(\xi_\perp^4))$} if only the {\it chaotic} (topologically non-trivial twisted) field fluctuations are regarded.}{5}
\glossaryentry{\glo{$l_{\rm s}$ ($l_{\rm l}$)}{characteristic distance a CR travels before leaving a flux tube in the CR source (loss) region (Eq. \ref{eq:llls})}}{7}
\glossaryentry{\glo{$L_{\rm s}= V_{\rm s}/A_{\rm s}$}{characteristic length of source}}{7}
\glossaryentry{\glo{$l_{\rm scatt}$}{CR pitch-angle scattering length $l_{\rm scatt} = v/\nu_\mu$}}{4}
\glossaryentry{\glo{$M_\|$}{number of inter phase flux tubes}}{8}
\glossaryentry{\glo{$m$}{particle mass}}{3}
\glossaryentry{\glo{$M_\perp$}{number of flux tubes forming the boundary}}{8}
\glossaryentry{\glo{$\mu$}{cosine of the CR pitch angle}}{3}
\glossaryentry{\glo{$N_{\rm s}$}{number of CR in source region}}{7}
\glossaryentry{\glo{$\nu_{\|}$}{CR escape frequency along flux tubes (Eqs. \ref{eq:escapetime}, \ref{eq:moreloss.2})}}{7}
\glossaryentry{\glo{$\nu_\mu$}{particle-wave pitch-angle scattering frequency}}{3}
\glossaryentry{\glo{$\nu_\perp$}{perpendicular escape frequency}}{9}
\glossaryentry{\glo{$\phi_B$}{magnetic flux}}{3}
\glossaryentry{\glo{$p$}{particle momentum}}{3}
\glossaryentry{\glo{$q$}{isotropic particle phase-space injection rate per magnetic flux (Eq. \ref{eq:qdef})}}{3}
\glossaryentry{\glo{$r_g$}{particle gyro-radius}}{3}
\glossaryentry{\glo{$r_{\rm i}$}{inner radius of (spherical) boundary layer}}{6}
\glossaryentry{\glo{$r_{\rm s}$}{source radius in spherical approximation}}{6}
\glossaryentry{\glo{$\sigma_\perp$}{fraction of flux tube surface in contact with loss region}}{8}
\glossaryentry{\glo{$\tau$}{particle escape time from a flux tube}}{3}
\glossaryentry{\glo{$\tau_{\rm s}$}{flux tube escape time in source region}}{7}
\glossaryentry{\glo{$\tau_{\rm T}$}{eddy turn-over time}}{6}
\glossaryentry{\glo{$v$}{particle velocity}}{3}
\glossaryentry{\glo{$V_{\rm s}$}{source volume}}{7}
\glossaryentry{\glo{$X_{\rm T}$}{increment factor of turbulence energy density}}{6}
\glossaryentry{\glo{$y = \lambda_\|/{l_{\rm s}}$}{field correlation length over particle diffusion length (Eq. \ref{eq:def_y})}}{8}
\glossaryentry{\glo{$Z$}{particle charge}}{3}
%\end{tabular}
\end{list}

\bibliographystyle{aabib99}

\end{document}